\documentclass[11pt,a4paper]{article}
\usepackage{amssymb,amsmath}
\usepackage{upgreek}
\usepackage[pdftex]{graphicx}
\usepackage[ruled,vlined, linesnumbered]{algorithm2e}
\usepackage{rotating}

\usepackage{color}
\usepackage{url}
\usepackage{subfigure}
\usepackage{xspace}

\urldef{\mailsbmi}\path|({mdeveci,kamer,umit}@bmi.osu.edu)|
\urldef{\mailsbu}\path|(bora.ucar@ens-lyon.fr)|

\newcommand{\GPUone}{GPUBFS\xspace}
\newcommand{\GPUtwo}{GPUBFS-WR\xspace}
\newcommand{\Ma}{{\cal M}}
\newcommand{\Pa}{{\cal P}}
\newcommand{\st}{{\sc L_0}\xspace}

\newcommand{\mcpf}{P-PFP\xspace}
\newcommand{\mchk}{P-HK\xspace}
\newcommand{\mcbfs}{P-DBFS\xspace}

\newcommand{\APSBFS}{APsB\xspace}
\newcommand{\APFBFS}{APFB\xspace}
\newcommand{\fixinconsistent}{{\sc FixMatching}\xspace}
\newcommand{\altswap}{{\sc{Alternate}}\xspace}

\newcommand{\supplementary} {\url{http://bmi.osu.edu/hpc/software/matchmaker2/maxCardMatch.html}}
\begin{document}


\title{GPU accelerated maximum cardinality matching algorithms for bipartite graphs}


\author{
Mehmet Deveci $^{1,2}$,
Kamer Kaya $^{1}$,
Bora U\c car $^{3}$,
\"{U}mit V. \c{C}ataly\"{u}rek $^{1,4}$ \\
  {$^1$ Dept. Biomedical Informatics, The Ohio State University}\\
  {$^2$ Dept. Computer Science and  Engineering, The Ohio State University}\\
  {$^3$ CNRS and LIP, ENS Lyon, France}\\
  {$^4$ Dept. Electrical and Computer Engineering, The Ohio State
    University}\\
  Email: {\textit{\{mdeveci,kamer,umit\}}@bmi.osu.edu and \textit{\{bora.ucar\}}@ens-lyon.fr}\\
}



\maketitle

\begin{abstract}
We design, implement, and evaluate GPU-based algorithms for the 
maximum cardinality matching problem in bipartite graphs.
Such algorithms 
have a variety of applications in computer science, scientific
computing, bioinformatics, and other areas.
To the best of our knowledge,
ours is the first study which focuses on GPU implementation of the maximum cardinality matching algorithms.
We compare
the proposed algorithms with serial and multicore implementations
from the literature on a large set of real-life problems where in
majority of the cases one of our GPU-accelerated algorithms is demonstrated to
be faster than both the sequential and multicore implementations.
\end{abstract}

\section{Introduction}
Bipartite graph matching is one of the fundamental problems in graph theory and combinatorial optimization.
The problem asks for a maximum set of vertex disjoint edges in a given bipartite graph.
It has many applications in a variety of fields such as image processing~\cite{kika:91}, chemical structure analysis~\cite{josz:95}, and
bioinformatics~\cite{azad2012multithreaded}~(see also another two discussed by Burkard~et~al.~\cite[Section 3.8]{budm:09}). Our motivating application lies in solving sparse linear 
systems of equations, as algorithms for computing a maximum cardinality bipartite matching are run routinely in the related solvers. In this setting, bipartite matching algorithms are used to see if the associated coefficient matrix is reducible; if so, substantial savings in computational requirements can be achieved~\cite[Chapter 6]{duer:86}.
 
Achieving good parallel performance on graph algorithms
is challenging, because they are memory bounded, and 
there are poor localities of the memory accesses. Moreover, because of 
the irregularity of the computations, it is difficult to exploit concurrency. 
Algorithms for the matching problem are no exception.
There have been recent studies that aim to improve the performance of 
matching algorithms on multicore and manycore architectures.
For example, Vasconcelos and Rosenhahn \cite{vasconcelos2009bipartite} propose
a GPU implementation of an algorithm for the maximum weighted matching problem on bipartite graphs. 
Fagginger Auer and Bisseling \cite{fagginger2012gpu} study 
an implementation of a greedy graph matching on GPU. 
\c{C}ataly\"{u}rek~et~al.~\cite{catalyurek2012multithreaded} propose
different greedy graph matching algorithms for multicore architectures. 
Azad~et~al.~\cite{ahrbka:12} 
introduce several multicore implementations of maximum cardinality matching 
algorithms on bipartite graphs. 

We propose GPU implementations of two maximum
cardinality matching algorithms. We analyze their
performance and employ further improvements. We thoroughly evaluate
their performance with a rigorous set of experiments on many bipartite
graphs from different applications. The experimental results
conclude that one of the proposed GPU-based implementation is faster than its
existing multicore counterparts.
 
The rest of this paper is organized as follows. The background
material, some related work, and a summary of contributions are presented in
Section~\ref{sec:back}. Section~\ref{sec:met} describes the proposed
GPU algorithms. The comparison of the proposed GPU-based
implementations with the existing sequential and multicore implementations
is given in
Section~\ref{sec:exp}. Section~\ref{sec:con} concludes the paper.

\vspace*{-1ex}
\section{Background and contributions}\label{sec:back}
\vspace*{-1ex}
A bipartite graph $G=(V_1\cup V_2,E)$ consists of a set of vertices $V_1\cup V_2$ where $V_1\cap V_2=\emptyset$, 
and a set of edges $E$ such that 
for each edge, one of the endpoints is in $V_1$ and other is in $V_2$.
Since our motivation lies in the sparse matrix domain, we will refer to the vertices in the two classes as row and column vertices.

A matching $\Ma$ in a graph $G$ is a subset of edges $E$ where a vertex in $V_1\cup V_2$ is in at most 
one edge in $\Ma$. 
Given a matching $\Ma$, a vertex $v$ is said to be matched by $\Ma$ if $v$ is in an edge of $\Ma$, otherwise $v$ is called unmatched.
The cardinality of a matching $\Ma$, denoted by $|\Ma|$, is the number of edges in $\Ma$. 
A matching $\Ma$ is called maximum, if no other matching $\Ma'$ with $|\Ma'|>|\Ma|$ exists.
For a matching $\Ma$, a path $\Pa$ in $G$ is called an $\Ma$-alternating if its edges are alternately in $\Ma$ and not in $\Ma$.
An $\Ma$-alternating path $\Pa$ is called $M$-augmenting if the start and end vertices of $\Pa$ are both unmatched.

There are three main classes of algorithms for finding the maximum
cardinality matchings in bipartite graphs. 
The first class of algorithms is based on augmenting paths~(see a detailed summary by Duff~et~al.~\cite{duff2011design}).
Push-relabel-based algorithms form a second class~\cite{Goldberg}. 
A third class, pseudoflow algorithms, is based on a more recent work~\cite{Hochbaum_pseudo}.
There are $\mathcal{O}(\sqrt{n}\tau)$ algorithms in the first two classes~(e.g., Hopcroft-Karp algorithm~\cite{hopcroft1973n} and a variant of the push relabel algorithm~\cite{goke:97}), where $n$ is the number of vertices and $\tau$ is the number of edges in the given bipartite graph.
This is asymptotically best bound for practical algorithms.
Most of the other known algorithms in the first two classes and in the third class have the running time complexity of  $\mathcal{O}(n\tau)$. 
These three classes of algorithms are described and compared in a recent study~\cite{klmu:12j}.
It has been demonstrated experimentally that the champions of the first two families 
are comparable in performance and better than that of the third family.
Since we investigate GPU acceleration of augmenting-path-based algorithms, a brief description of them is given below (the reader is invited to two
recent papers~\cite{duff2011design,klmu:12j} and the original resources cited in those papers for other algorithms).

Algorithms based on augmenting paths follow the following common pattern. Given an initial matching $\Ma$ (possible empty), they search for an $\Ma$-augmenting
path $\Pa$. If none exists, then the
matching $\Ma$ is maximum by a theorem of Berge~\cite{berg:57}. 
Otherwise, the augmenting path $\Pa$ is used to
increase the cardinality of $\Ma$ by setting $\Ma = \Ma \oplus E(\Pa)$
where $E(\Pa)$ is the edge set of the path $\Pa$, and $\Ma \oplus E(\Pa) = (\Ma \cup E(\Pa)) \setminus (\Ma \cap E(\Pa))$ is the symmetric difference. This inverts the membership in $\Ma$ for all edges
of $\Pa$. Since both the first and the last edge of $\Pa$ were unmatched in $\Ma$, we have
$|\Ma \oplus E(\Pa)|=|\Ma|+1$.
The augmenting-path-based algorithms differ in the way these augmenting paths are found and the associated augmentations are realized.
They mainly use either breadth-first-search (BFS), or depth-first-search (DFS), or combination 
of these two techniques to locate and perform the augmenting paths. 

Multicore 
counterparts of a number of augmenting-path based algorithms are proposed in a recent work~\cite{ahrbka:12}. The parallelization 
of these algorithms is achieved by using atomic operations at BFS and/or DFS steps 
of the algorithm. Although using atomic operations might not harm the performance 
on a multicore machine, they should be avoided in a GPU 
implementation because of very large number of concurrent thread executions. 

As a reasonably efficient DFS is not feasible with GPUs, we accelerate two BFS-based algorithms, called HK~\cite{hopcroft1973n} and HKDW~\cite{duwi:88}.
HK has the best known worst-case running time complexity of $O(\sqrt{n}\tau)$ for a bipartite graph with $n$ vertices and $\tau$ edges.
HKDW is a variant of HK and incorporates techniques to improve the practical running time while having the same worst-case time complexity.
Both of these algorithms use BFS to locate the shortest augmenting paths from unmatched columns, 
and then use DFS-based searches restricted to a certain part of the input graph to augment along a maximal set of disjoint augmenting paths.
HKDW performs another set of DFS-based searches to augment using the remaining unmatched rows. 
As is clear, the DFS-based searches will be a big obstacle to achieve efficiency. 
In order to overcome this hurdle, we propose a scheme which alternates the edges of a number of 
augmenting paths with a parallel scheme that resembles to a breadth expansion in BFS. 
The proposed scheme offers a high degree of parallelism but does not guarantee a maximal set of augmentations, potentially increasing the worst case time complexity to  $O(n\tau)$ on a sequential machine. 
In other words, we trade theoretical worst case time complexity with a higher degree of parallelism to achieve better practical running time with a GPU. 

\section{Methods}\label{sec:met}
We propose two algorithms for the GPU implementation of maximum cardinality 
matching. 
These algorithms use BFS to find augmenting paths, speculatively perform some of them, and
fix any inconsistencies that can be resulting from speculative augmentations. 

The overall structure of the first GPU-based algorithm is given in Algorithm~\ref{alg:apbfsee},  \APSBFS.
It largely follows the common structure of most of the existing sequential algorithms, and corresponds to HK. 
It performs a combined BFS starting from all unmatched columns to find unmatched rows, thus locating augmenting paths.
Some of those augmentations are then realized using a function called \altswap (will be described later).
The parallelism is exploited inside the {\sc InitBfsArray}, {\sc BFS}, \altswap, and \fixinconsistent functions. 
Algorithm~\ref{alg:apbfsee} is given the adjacency list 
of the bipartite graph with its number of rows and columns. 
Any prior matching 
is given in $rmatch$ and $cmatch$ arrays as follows: $rmatch[r] = c$ and
$cmatch[c] = r$, if the row $r$ is matched to the column $c$;  
$rmatch[r] = -1$, if $r$ is unmatched; $cmatch[c] = -1$, if $c$ is unmatched.

\begin{algorithm}
  \small
  \caption{{\sc Shortest augmenting paths (\APSBFS)} }
  \label{alg:apbfsee}
  \KwData{$cxadj, cadj, nc, nr, rmatch, cmatch$}
  
$augmenting\_path\_found \gets {\bf true}$;\\
\While{$augmenting\_path\_found$}{
  $bfs\_level \gets \st{}$;\\
  {\sc InitBfsArray}$(bfs\_array, cmatch,\st{})$;\\	
  $vertex\_inserted \gets {\bf true}$;\\
  \While{$vertex\_inserted $}{
	$predecessor \gets  ${\sc Bfs}$(bfs\_level, bfs\_array, cxadj, cadj, nc, rmatch,$\\
		\hspace*{21ex}$ vertex\_inserted, augmenting\_path\_found)$;\\
	\If{$augmenting\_path\_found$}{
		$break$;
	}
	
	$bfs\_level \gets bfs\_level + 1$;
  }
  
  $\langle cmatch, rmatch\rangle \gets$ \altswap$(cmatch, rmatch, nc,  predecessor)$; \\
  $\langle cmatch, rmatch\rangle \gets$ \fixinconsistent$(cmatch, rmatch)$; \\
}
\end{algorithm}

%
%
%
%

The outer loop of Algorithm~\ref{alg:apbfsee} iterates until no more augmenting paths are
found, thereby guaranteeing a maximum matching.
The inner loop is responsible from completing the breadth-first-search
of the augmenting paths. A single iteration of this loop corresponds to a level of BFS. The inner loop 
iterates until all shortest augmenting paths are found. Then, the edges in 
these shortest augmenting paths are alternated inside \altswap function. 
Unlike the sequential HK algorithm, \APSBFS does not find a maximal set of augmenting paths. 

By removing the lines 9 and 10 of Algorithm~\ref{alg:apbfsee}, another matching algorithm is obtained. 
This method will continue with the BFSs until all possible
unmatched rows are found; it can be therefore considered as the GPU 
implementation of the HKDW algorithm. This variant is called \APFBFS{}.

\begin{algorithm}
  \small 
  \caption{{\sc BFS Kernel Function-1 (\GPUone{})} }
  \label{alg:kernelbfs}
  \KwData{$bfs\_level, bfs\_array,
		cxadj, cadj, nc, rmatch, $\\
		\hspace*{10ex}$vertex\_inserted,
		augmenting\_path\_found$} 

	$process\_cnt \gets {\sc getProcessCount} (nc)$;\\
	  \For {$i$ {\bf from}  $0$ {\bf to} $process\_cnt - 1$}{
  		$col\_vertex \gets i \times tot\_thread\_num + tid$;\\
		\If{$bfs\_array[col\_vertex]$ $=$ $bfs\_level$}{
			\For {$j$ {\bf from}  $cxadj[col\_vertex]$ {\bf to} $cxadj[col\_vertex + 1]$ }{
				$neighbor\_row \gets cadj[j]$;\\
				$col\_match \gets rmatch[neighbor\_row]$;\\
				\If{$col\_match > -1$}{
					\If{$bfs\_array[col\_match]$ $=$ $\st{}$ $-1$ }{
						$vertex\_inserted \gets {\bf true}$;\\
						$bfs\_array[col\_match] \gets bfs\_level + 1$;\\
						$predeccesor[neighbor\_row] \gets col\_vertex$;\\
					}
				}
				\Else{
					 \If{$col\_match $=$ -1$}{
						$rmatch[neighbor\_row] \gets -2 $;\\			
						$predeccesor[neighbor\_row] \gets col\_vertex$;\\
						$augmenting\_path\_found \gets {\bf true}$;\\
					 }
				}
			}
		}
  	}

\end{algorithm}

We propose two implementations of the BFS kernel. 
Algorithm~\ref{alg:kernelbfs} is the first one. The BFS kernel is 
responsible from a single level BFS expansion. 
That is, it takes the set of vertices at a BFS level and adds the union of the unvisited neighbors of
those vertices as the next level of vertices.
Initially, the input $bfs\_array$ filled with $bfs\_array[c] = \st{} - 1$ if $cmatch[c] > -1$ and $bfs\_array[c] = \st{}$ if 
$cmatch[c] = -1$ by a simple {\sc InitBfsArray} kernel ($\st{}$ denotes BFS start level).

 The GPU threads partition the columns vertices in a single dimension. Each thread with id $tid$
 is assigned a number of columns which is obtained 
 via the following function:
 
\resizebox{0.95\columnwidth}{!}{%
$
getProcessCount(nc) = \left\{ 
  \begin{array}{l l}
    \lceil \frac{nc} {tot\_thread\_num}  \rceil&  \text{if $tid < nc \bmod tot\_thread\_num$,}\\
    \lfloor \frac{nc} {tot\_thread\_num}  \rfloor &  \text{otherwise.}
  \end{array} \right.
  $
 \noindent
 }
 
 Once the number of columns are obtained, the threads traverse their 
 first assigned column vertex. The indices of the columns assigned to a thread 
 differ by $tot\_thread\_num$ to allow coalesced global memory accesses. 
 Threads traverse the neighboring row vertices of the current column, 
 if its BFS level is equal to the current $bfs\_level$. If a thread encounters
 a matched row during the traversal, its matching column is retrieved. 
 If the column is not traversed yet, its $bfs\_level$ is marked on $bfs\_array$. 
 On the other hand, when a thread encounters an unmatched row, 
 an augmenting path is found. In this case, the match of the neighbor 
 row is set to $-2$, and this information is used by \altswap later.

\begin{algorithm}
  \small 
  \caption{\altswap}
  \label{alg:kernelswap_edges}
  \KwData{$
		cmatch, rmatch, nc, nr , predecessor$} 

	$process\_vcnt \gets {\sc getProcessCount} (nr)$;\\
	  \For {$i$ {\bf from}  $0$ {\bf to} $process\_vcnt - 1$}{
  		$row\_vertex \gets i \times tot\_thread\_num + tid$;\\
		\If{$rmatch[row\_vertex]$ $=$ $-2$}{
			\While{$row\_vertex \neq -1$}{
				$matched\_col \gets predecessor[row\_vertex]$;\\
				$matched\_row \gets cmatch[matched\_col]$ ;\\
				\If{predecessor[matched\_row] $=$ matched\_col}{
					$break$;\\
				}	
				$cmatch[matched\_col] \gets row\_vertex$;\\
				$rmatch[row\_vertex] \gets matched\_col$;\\
				$row\_vertex \gets matched\_row$;\\
				
			}
			
		}
  	}
\end{algorithm}

\begin{figure} [!htb]
\center
\includegraphics[width=0.40\textwidth]{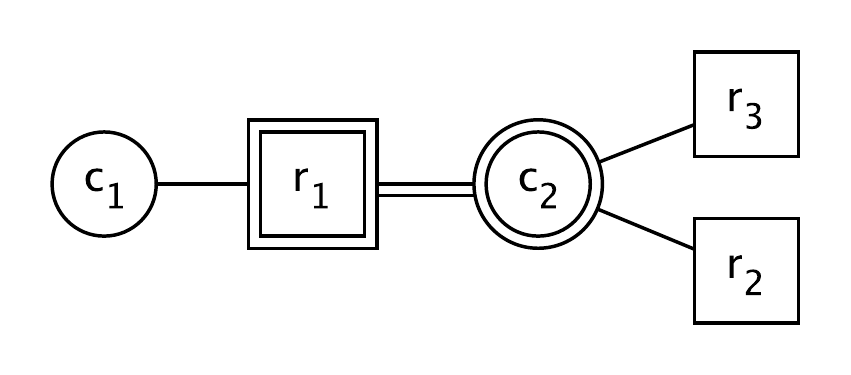}
\vspace*{-1ex}
\caption{Vertices $r_1$ and $c_2$ are matched; others are not.
Two augmenting paths starting from $c_1$ are possible.}

 \label{fig:swapedges}
\end{figure}

Algorithm~\ref{alg:kernelswap_edges} gives the description of the \altswap function. This kernel 
alternates the matching edges with unmatching edges of the augmenting paths found; some of those paths end up being augmenting ones and some are only partially alternated.
Here, each thread is assigned a number of rows. Since $rmatch$ of an unmatched row 
(that is also an endpoint of an augmenting path) has been set to $-2$ in the BFS kernel, 
only the threads whose row vertex's match
is $-2$ start \altswap. Since there might be 
several augmenting paths for an unmatched
column, race conditions while writing on
$cmatch$ and $rmatch$ arrays are possible. Such a race condition might cause
infinite loops (inner while loop) or inconsistencies, if care is not taken.  
We prevent these by checking the predecessor of a matched row 
(line-8). For example, in Fig.~\ref{fig:swapedges},
two different
augmenting paths that end with $r_2$ and $r_3$ are found for $c_1$. If the thread
of $r_2$ starts before the thread of $r_3$ in \altswap, the match of $c_2$ will be updated
to $r_2$ (line-10). Then, $r_3$'s thread will read $matched\_row$ of $c_2$
as $r_2$ (line-7). This would cause an infinite loop
without the check at line-8. Inconsistencies may occur when the
threads of $r_2$ and $r_3$ are in the same warp. In this case, 
the if-check will not hold for both threads, and their row vertices will
be written on $cmatch$ (line-10).
Since only one thread will be successful at writing, 
this will cause an inconsistency. Such inconsistencies are fixed
by \fixinconsistent{} kernel which implements: $\mbox {$rmatch[r] \gets -1$}$ for any $r$ satisfying $cmatch[rmatch[r]] \neq r$.

Algorithm~\ref{alg:kernelbfs2} gives the description of a slightly different BFS kernel function.  
This function takes $root$ array as an extra argument. Initially, the root array
is filled with $root[c] = 0$ if $cmatch[c] > -1$, and $root[c] = c$ if $cmatch[c] = -1$.
This array holds the root (as the index of the column vertex) of an augmenting path, and this information is 
transferred down during BFS. Whenever an augmenting path is found, the entry in 
$bfs\_array$ for the root of the augmenting path is set to $\st{}-2$. 
This information is used at the beginning of BFS kernel. No more BFS traversals is done,
if an augmenting path is found for the root of the traversed column vertex. Therefore,
while the method increases the global memory accesses by introducing 
an extra array, it provides an early exit mechanism for BFS. 

\begin{algorithm}
  \small 
  \caption{{\sc BFS Kernel Function-2 (\GPUtwo{})} }
  \label{alg:kernelbfs2}
  \KwData{$bfs\_level, bfs\_array,
		cxadj, cadj, nc, rmatch, root$\\
		\hspace*{10ex}$vertex\_inserted,
		augmenting\_path\_found$} 

	$process\_cnt \gets {\sc getProcessCount} (nc)$;\\
	  \For {$i$ {\bf from}  $0$ {\bf to} $process\_cnt - 1$}{
  		$col\_vertex \gets i \times tot\_thread\_num + tid$;\\
		\If{$bfs\_array[col\_vertex]$ $=$ $bfs\_level$}{
			$myRoot \gets root[col\_vertex]$;\\
			\If {$bfs\_array[myRoot] < \st{} - 1$}{
				$continue$;\\
			}
			\For {$j$ {\bf from}  $cxadj[col\_vertex]$ {\bf to} $cxadj[col\_vertex + 1]$ }{
				$neighbor\_row \gets cadj[j]$;\\
				$col\_match \gets rmatch[neighbor\_row]$;\\
				\If{$col\_match > -1$}{
					\If{$bfs\_array[col\_match]$ $=$  $\st{}$ $-1$}{
						$vertex\_inserted \gets {\bf true}$;\\
						$bfs\_array[col\_match] \gets bfs\_level + 1$;\\
						$root [col\_match] \gets myRoot$;\\
						$predeccesor[neighbor\_row] \gets col\_vertex$;\\
					}
				}
				\Else{
					 \If{$col\_match $=$ -1$}{
					 	$bfs\_array[myRoot] \gets \st{} - 2$;\\
						$rmatch[neighbor\_row] \gets -2 $;\\			
						$predeccesor[neighbor\_row] \gets col\_vertex$;\\
						$augmenting\_path\_found \gets {\bf true}$;\\
					 }
				}
			}
		}
  	}
\vspace*{-1ex}

\end{algorithm}

We further improve \GPUtwo{} by making use of the arrays $root$ and $bfs\_array$. 
BFS kernels might find several rows to match with the same 
unmatched column, and set $rmatch[\cdot]$ to $-2$ for each.
These cause \altswap to start
from several rows that can be matched with the same unmatched column.
Therefore, it may perform unnecessary alternations, until these augmenting
paths intersect. Conflicts may occur at these intersection points (which are then resolved  
with \fixinconsistent function).
By choosing $\st{}$ as 2, we can limit the range of the values that $bfs\_array$ takes to positive 
numbers. Therefore, by setting the $bfs\_array$ to $-(neighbor\_row)$ at line 19 of 
Algorithm~\ref{alg:kernelbfs2},  
we can provide more 
information to the \altswap function. With this, \altswap can determine
the beginning and the end of an augmenting path, and it can alternate only among
the correct augmenting paths.  \APSBFS-\GPUtwo{}~(and \altswap function used
together) includes these improvements. However, they are not included in
\APFBFS-\GPUtwo{} since they do not improve its performance.

\vspace*{-3ex}
\section{Experiments}\label{sec:exp}
\vspace*{-1ex}
The running time of the proposed implementations are compared against the 
sequential HK and PFP implementations~\cite{duff2011design}, and
against the multicore parallel implementations 
\mcpf, \mcbfs, and \mchk~\cite{ahrbka:12}. The 
CPU implementations are tested on a computer with 2.27GHz dual quad-core Intel Xeon CPUs with 
2-way hyper-threading and 48GB main memory. The algorithms are implemented in 
C++ and OpenMP. The GPU implementations are tested
on NVIDIA Tesla C2050 with usable 2.6GB of global memory. 
C2050 is equipped with 14 multiprocessors each containing 32 CUDA cores, 
totaling 448 CUDA cores. The implementations are compiled with gcc-4.4.4, cuda-4.2.9 and -O2 optimization flag. For the multicore algorithms, 
8 threads are used.  
A standard heuristic (called the cheap matching, see~\cite{duff2011design}) is used to initialize all tested algorithms.
We compare the running
time of the matching algorithms after this common initialization.

Two different main algorithms \APFBFS and \APSBFS can use two different BFS kernel functions (\GPUone{} and \GPUtwo{}).
Moreover, each of these algorithms can have two 
versions (i) CT: uses a constant number of threads with fixed number of grid and block size
($256\times 256$) and assigns multiple vertices to each thread; (ii) MT: tries to assign one vertex to each thread. The number
of threads used in the second version is chosen as $\mbox{MT} = \min(nc, \#threads)$
where $nc$ is the number of columns, and $\#threads$ is the 
maximum number of threads of the architecture. 
Therefore, we have eight GPU-based algorithms.

The algorithms are run on bipartite graphs corresponding to 70 different matrices 
from variety of classes 
at UFL matrix collection~\cite{dahu:11}. 
We also permuted the matrices randomly by rows and columns and included them as a second set (labeled RCP). These permutations usually render the problems
harder for the augmenting-path-based algorithms~\cite{duff2011design}. 
For both sets, we report the performance for a smaller subset which contains those matrices in which 
at least one of the sequential algorithms took more than one second.
We call these sets O\_S1 (28 matrices) and RCP\_S1 (50 matrices). We also have another two subsets called O\_Hardest20 and
RCP\_Hardest20 that contain the set of 20 matrices on which the sequential algorithms required the longest running time.

\begin{table}
\caption{Geometric mean of the running time of the GPU algorithms on different sets of instances.
}
\vspace*{-2ex}
\label{tab:gputime}
\center
\resizebox{\columnwidth}{!}{%
\begin{tabular}{l|r|r|r|r|r|r|r|r}
\multicolumn{1}{c|}{} & \multicolumn{4}{|c|}{\APFBFS{}} & \multicolumn{4}{|c}{\APSBFS{}} \\ \cline{2-9}

\multicolumn{1}{c|}{} & \multicolumn{2}{|c|}{\GPUone} & \multicolumn{2}{|c|}{\GPUtwo}  & \multicolumn{2}{|c|}{\GPUone}  &
\multicolumn{2}{c}{\GPUtwo} \\  \cline{2-9}
 & MT & CT & MT& CT & MT & CT & MT & CT\\\hline
O\_S1 & 2.96 & 1.89 & 2.12 & {\bf 1.34} & 3.68 & 2.88 & 2.98 & 2.27\\\hline
O\_Hardest20 & 4.28 & 2.70 & 3.21 & {\bf 1.93} & 5.23 & 4.14 & 4.20 & 3.13\\\hline
RCP\_S1 & 3.66 & 3.24 & 1.13 & {\bf 1.05} & 3.52 & 3.33 & 2.22 & 2.14\\\hline
RCP\_Hardest20 & 7.27 & 5.79 & 3.37 & {\bf 2.85} & 12.06 & 10.75 & 8.17 & 7.41\\\hline
\end{tabular}
}
\end{table}

\begin{figure} [!htb]

\subfigure[Hamrle3]{
\includegraphics[width=0.45\textwidth]{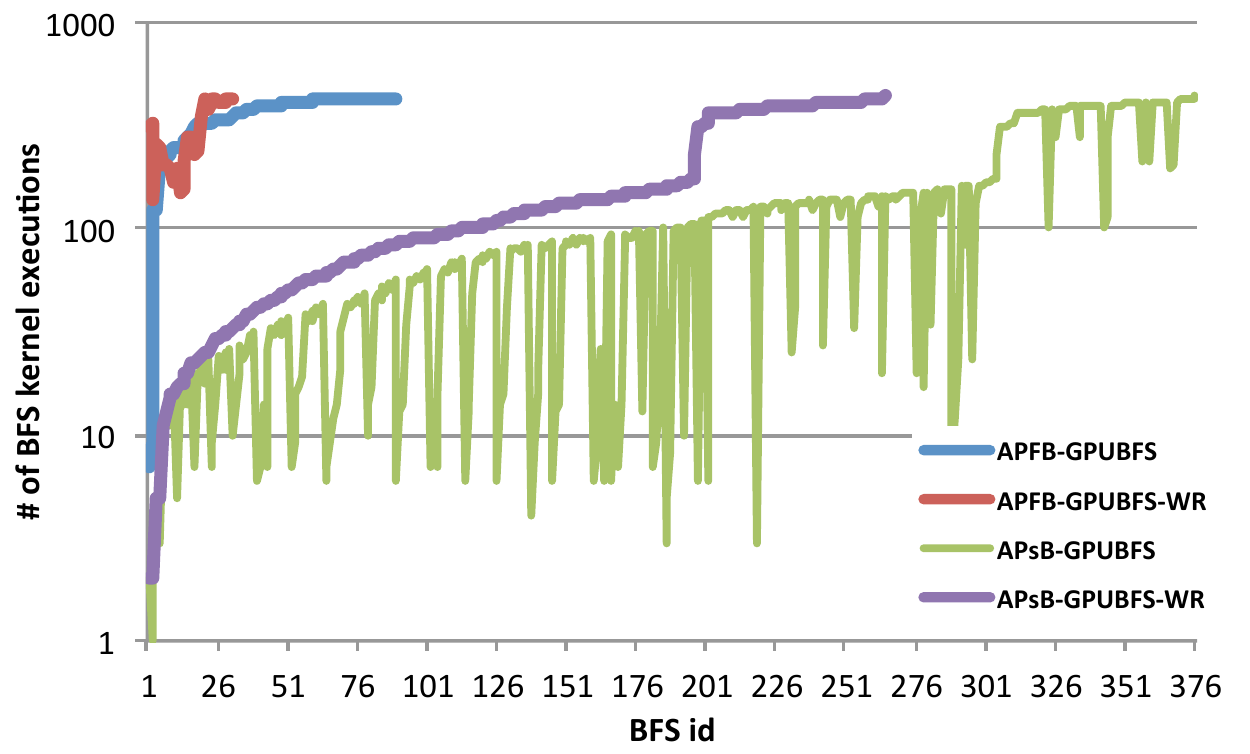}
 \label{fig:bfs_ham}
}
\subfigure[Delanuay\_n23]{
\includegraphics[width=0.45\textwidth]{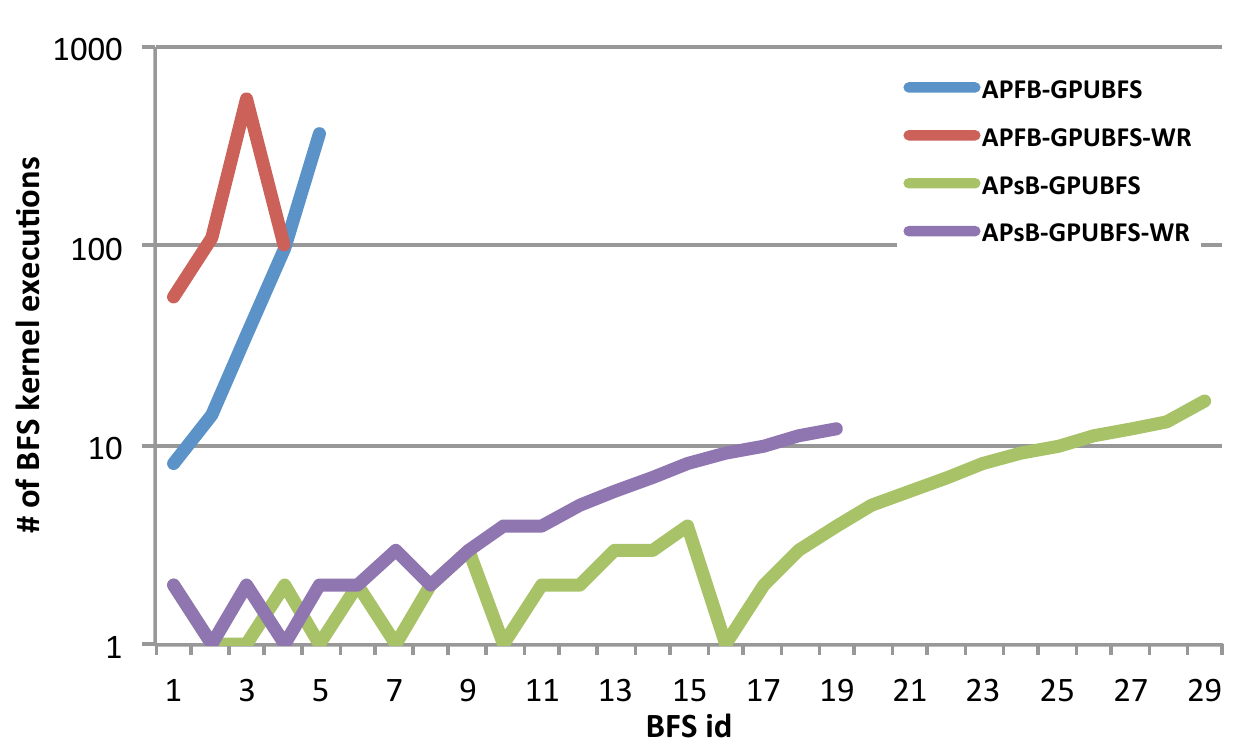}
 \label{fig:bfs_del}
}
\vspace*{-1em}
\caption{
The BFS ids and the number of kernel executions for each BFS in \APSBFS and \APFBFS variants for two graphs. 
The $x$ axis shows the id of the {\bf while} iteration at line~2 of \APSBFS.
The $y$ axis shows the number of the {\bf while} iterations at line~6 of \APSBFS.
}
\vspace*{-4ex}
\end{figure}

First, we compare the performance of the proposed GPU algorithms. Table~\ref{tab:gputime} 
shows the geometric mean of the running time on different sets. As we see from the table, 
using constant number of threads (CT) always increases the performance of an algorithm, since it
increases the granularity of the work performed by each thread. \GPUtwo{} is always
faster than \GPUone{}. This is because of the unnecessary BFS traversals in the \GPUone{}
algorithm. \GPUone{} cannot determine whether an augmenting path has already 
been found for an unmatched column, therefore it will continue to explore.
This unnecessary BFS traversals not only increase the BFS time, but also reduce the likelihood of
finding an augmenting path for other unmatched columns. 
Moreover, the \altswap scheme turns out to be more suitable for \APFBFS than \APSBFS, 
in which case it can augment along more paths (there is a larger 
set of possibilities). For example, Figs.~\ref{fig:bfs_ham} and~\ref{fig:bfs_del} show the number 
of BFS iterations and the number of BFS levels in each iteration for, respectively, Hamrle3 and Delanuay\_n23 graphs. 
As clearly seen from both of the figures, 
\APFBFS variants converges in smaller number of iterations than \APSBFS variants; and for most of the graphs,
the total number of BFS kernel calls are less for \APFBFS (as in Fig.~\ref{fig:bfs_ham}). 
However, for a small subset of the graphs, although the augmenting path exploration
of \APSBFS converges in larger number of iterations, the numbers of the BFS levels in each iterations
are much less than \APFBFS (as in Fig.~\ref{fig:bfs_del}). Unlike the general case,
\APSBFS outperforms \APFBFS in such cases.
Since \APFBFS using \GPUtwo{}   
and CT 
almost always obtains the best performance, we only compare the performance of this 
algorithm with other implementations in the following.

\begin{figure} [!htb]
\subfigure[Original graphs]{
\includegraphics[width=0.45\textwidth]{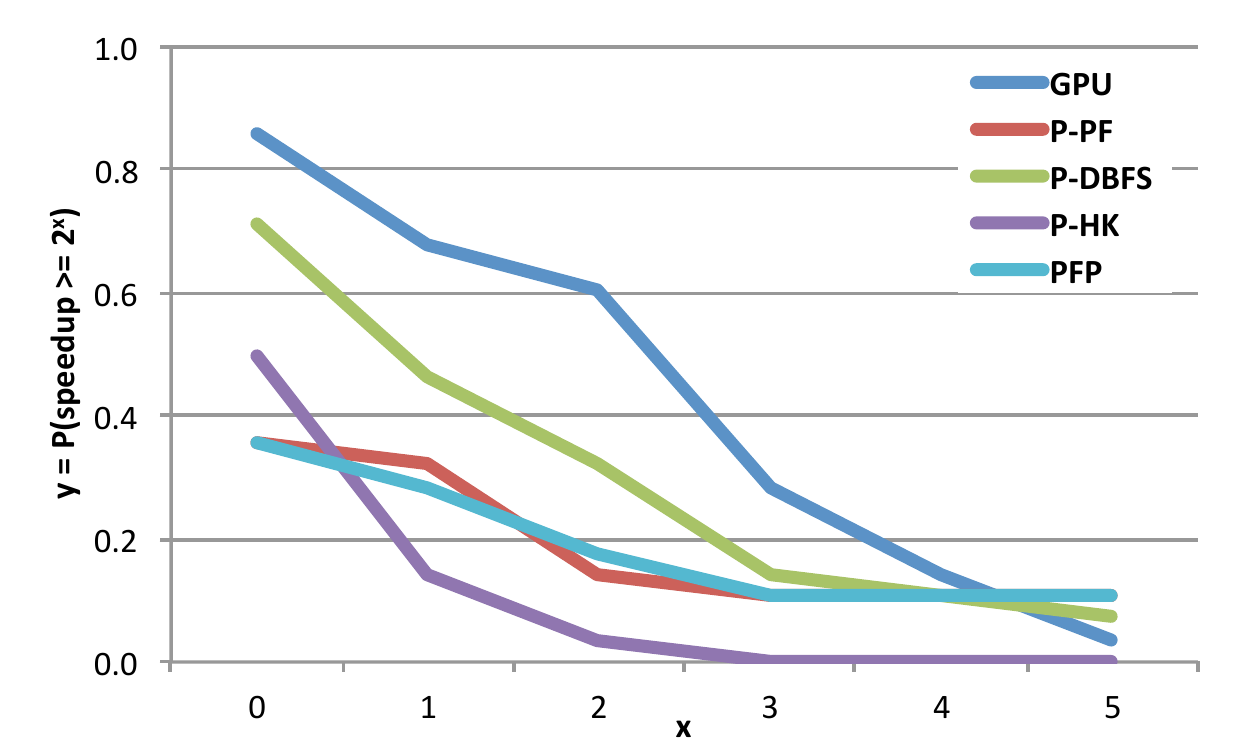}
 \label{fig:subos}
}
\subfigure[Permuted graphs]{
\includegraphics[width=0.45\textwidth]{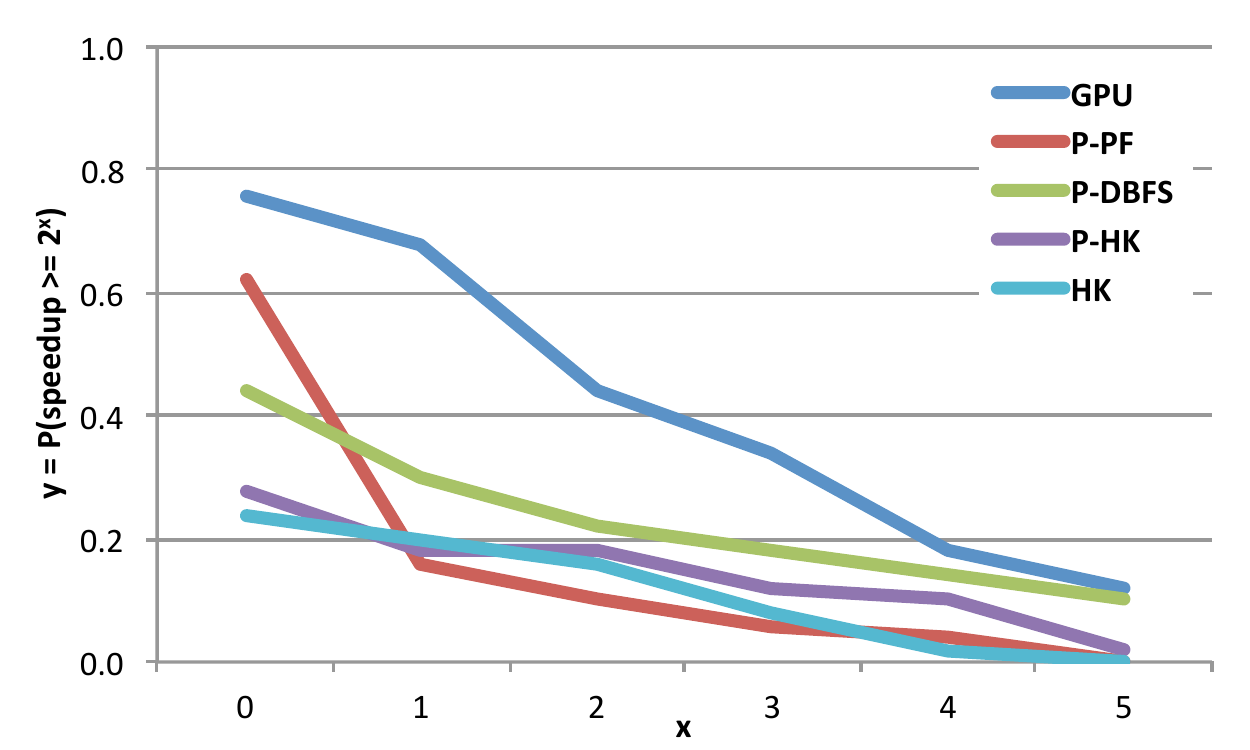}
 \label{fig:subrcs}
}
\vspace*{-1em}
\caption{Log-scaled speedup profiles.}
\end{figure}

Figures~\ref{fig:subos} and~\ref{fig:subrcs} give the log-scaled 
speedup profiles of the best GPU and multicore algorithms on 
the original and permuted graphs. The speedups are calculated with respect to the fastest of
the sequential algorithms PFP and HK (on the original graphs HK was faster; on the permuted ones PFP was faster). 
A point $(x,y)$ in the plots corresponds to the 
probability of obtaining at least $2^x$ speedup is $y$. As the plots show, the GPU 
algorithm has the best overall speedup. It is faster than the sequential HK 
algorithm for $86 \%$ of the original graphs, while it is faster than PFP on $76 \%$
of the permuted graphs. \mcbfs obtains the best performance among the multicore algorithms.
However, its performance degrades on permuted graphs.
Although \mcpf is more robust than \mcbfs to permutations, its overall performance
is inferior to that of \mcbfs. \mchk is outperformed by the other algorithms in both sets.

\begin{figure} [!htb]
\subfigure[Original graphs]{
\includegraphics[width=0.45\textwidth]{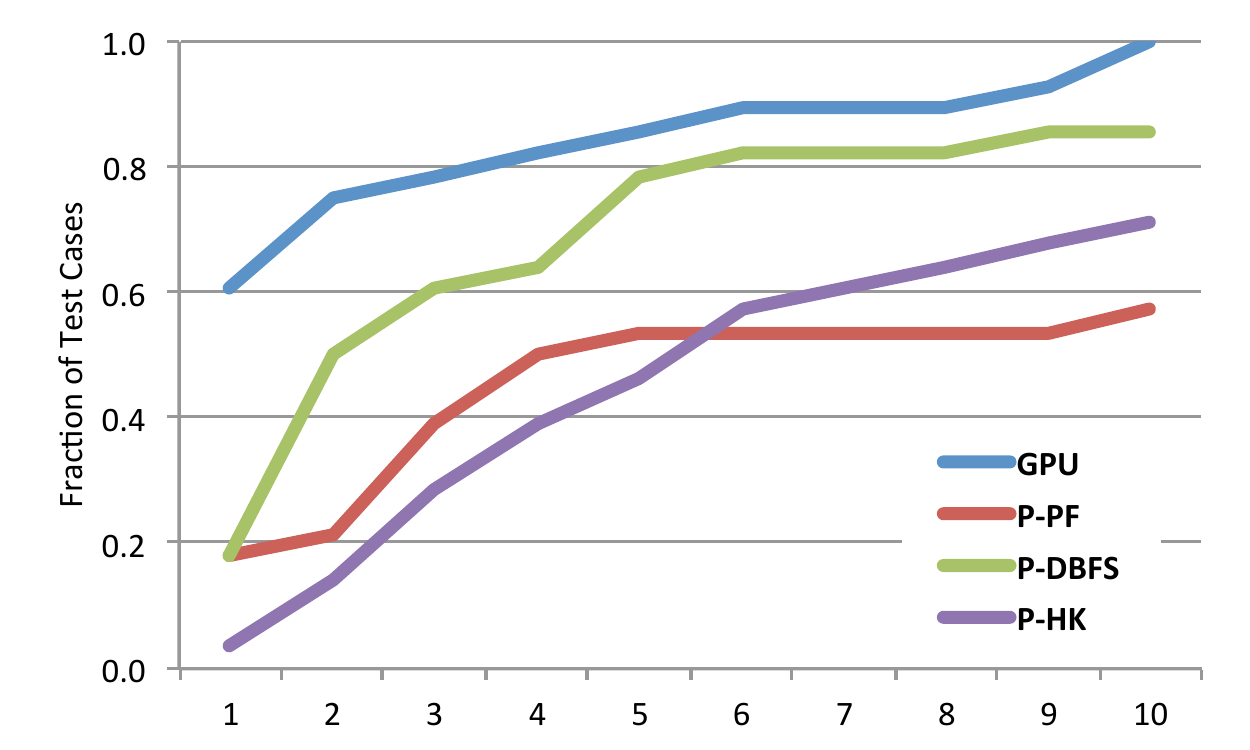}
 \label{fig:subop}
}
\subfigure[Permuted graphs]{
\includegraphics[width=0.45\textwidth]{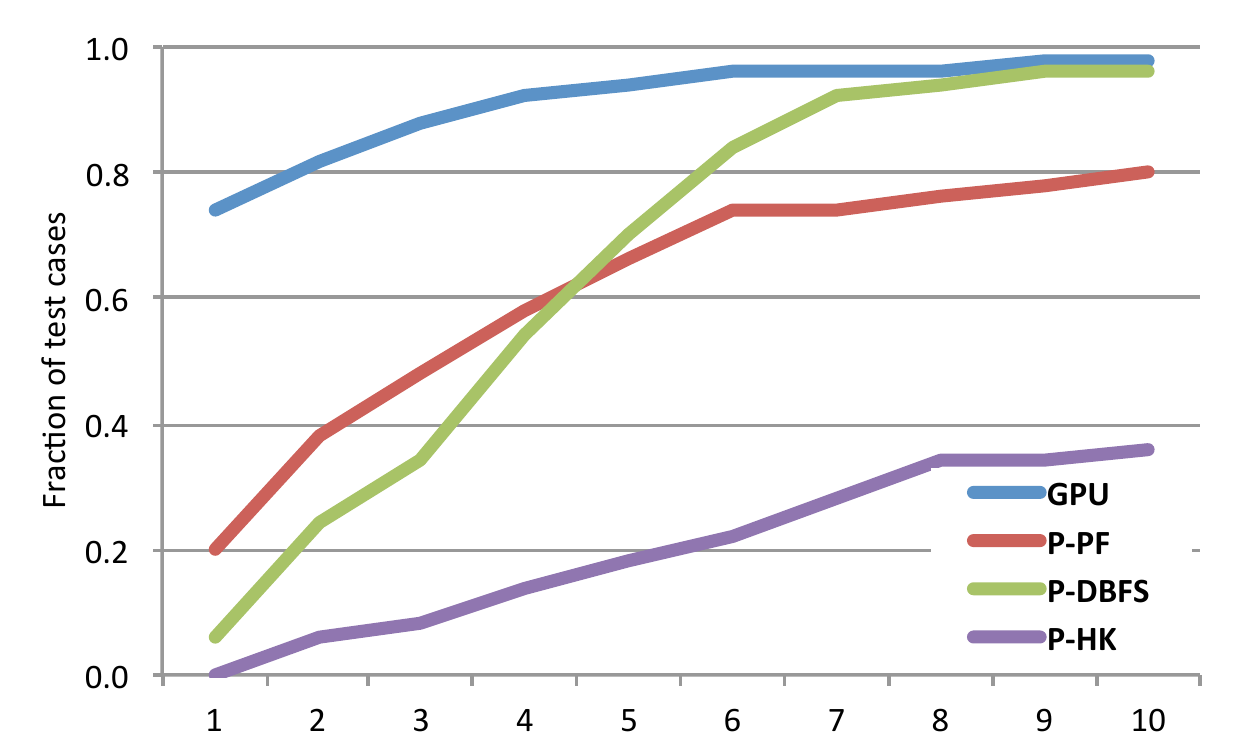}
 \label{fig:subrcp}
}
\vspace*{-1em}
\caption{Performance profiles}

\end{figure}

Figures~\ref{fig:subop} and~\ref{fig:subrcp} show the performance profiles of 
the GPU and multicore algorithms. A point $(x,y)$ in this plot means that with $y$ probability, 
the algorithm obtains a performance that is at most $x$ times worse than the best 
running time. The plots clearly show the separation among the GPU algorithm and the multicore ones, especially for original graphs and for $x\leq 7$ for the permuted ones, thus marking GPU as the fastest in most cases.
In particular, GPU algorithm obtains the 
best performance in $61 \%$ of the original graphs, while this ratio increases to $74\%$ for the permuted 
ones.

\begin{figure} [!htb]
\center
\includegraphics[width=0.70\textwidth]{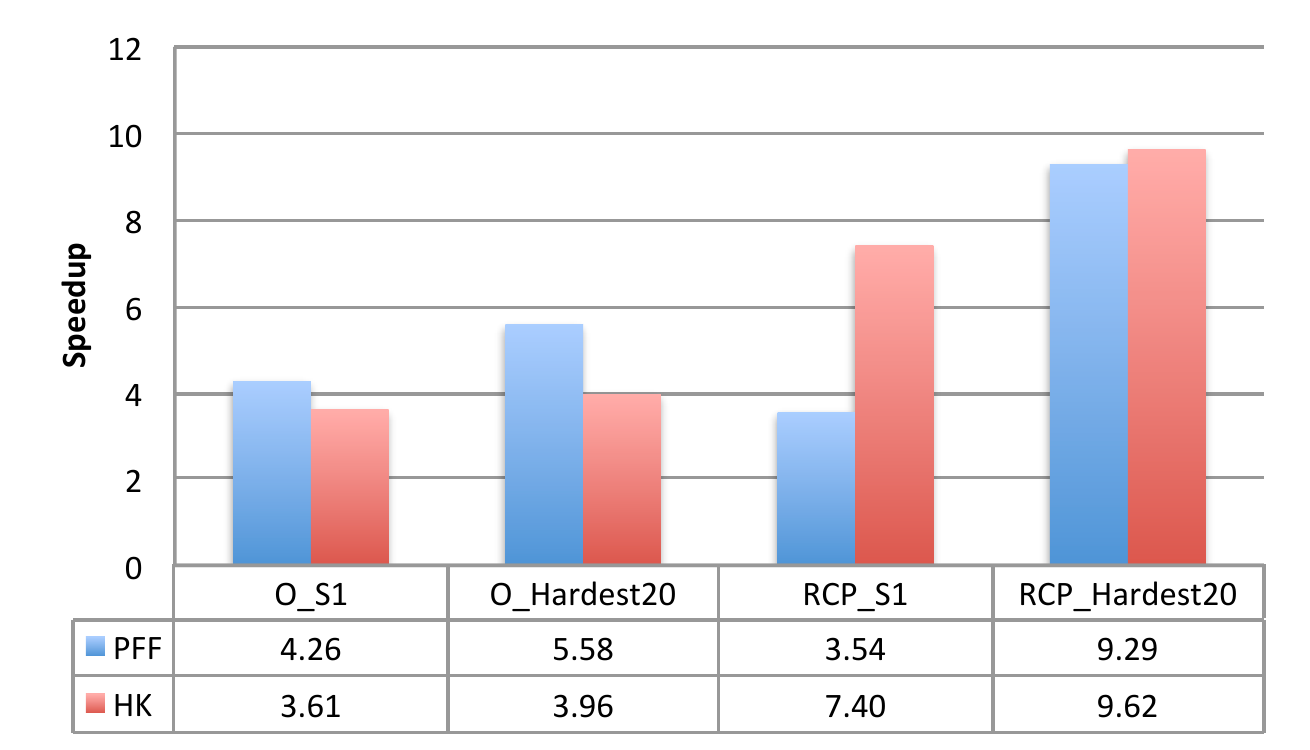}
\vspace*{-2ex}
\caption{Overall speedup of the proposed GPU algorithm w.r.t. PFP (left bars) and HK (right bars) algorithms. 
}

 \label{fig:overallspeedup}
\end{figure}

Figure~\ref{fig:overallspeedup} gives the overall speedups. The proposed GPU
algorithm obtains average speedup values of at least $3.61$ and $3.54$ on, respectively, original and permuted graphs.
The speedups increase for the hardest instances, where the GPU algorithm achieves $3.96$ and $9.29$ speedup, respectively, on original and permuted graphs. Table~\ref{tab:actualrun}
gives the actual running time for O\_Hardest20 sets for the best GPU and multicore algorithms, 
together with the sequential algorithms (the running time of all mentioned algorithms on 
the complete set of graphs can be found at \supplementary).
As seen from this table, except six instances among the original graphs and another two among the permuted graphs, the GPU algorithm is faster than the best sequential algorithm. It is also faster than the multicore ones in all, except five original graphs. 

\begin{table}[!htb]
\caption{The actual running time of each algorithm for the O\_Hardest20 set.}
\vspace*{-4ex}
\label{tab:actualrun}
\center
\resizebox{\columnwidth}{!}{%
\begin{tabular}{l|r|r|r|r||r|r|r|r}
& \multicolumn{4}{|c||}{Original graphs}& \multicolumn{4}{|c}{Permuted graphs}\\ \cline{2-9}
Matrix name & GPU & \mcbfs & PFP & HK & GPU & \mcbfs & PFP & HK\\\hline
roadNet-CA & {\bf 0.34} & 0.53 & 0.95 & 2.48 & {\bf 0.39} & 1.88 & 3.05 & 4.89\\\hline
delaunay\_n23 & {\bf 0.96} & 1.26 & 2.68 & 1.11 & {\bf 0.90} & 5.56 & 3.27 & 14.34\\\hline
coPapersDBLP & {\bf 0.42} & 6.27 & 3.11 & 1.62 & 0.38 & 1.25 & {\bf 0.29} & 1.26\\\hline
kron\_g500-logn21 & {\bf 0.99} & 1.50 & 5.37 & 4.73 & {\bf 3.71} & 4.01 & 64.29 & 16.08\\\hline
amazon-2008 & {\bf 0.11} & 0.18 & 6.11 & 1.85 & {\bf 0.41} & 1.37 & 61.32 & 4.69\\\hline
delaunay\_n24 & {\bf 1.98} & 2.41 & 6.43 & 2.22 & {\bf 1.86} & 12.84 & 6.92 & 35.24\\\hline
as-Skitter & {\bf 0.49} & 1.89 & 7.79 & 3.56 & {\bf 3.27} & 5.74 & 472.63 & 29.63\\\hline
amazon0505 & {\bf 0.18} & 22.70 & 9.05 & 1.87 & {\bf 0.24} & 15.23 & 17.59 & 2.23\\\hline
wikipedia-20070206 & {\bf 1.09} & 5.24 & 11.98 & 6.52 & {\bf 1.05} & 5.99 & 9.74 & 5.73\\\hline
Hamrle3 & 1.36 & 2.70 & {\bf 0.04} & 12.61 & {\bf 3.85} & 7.39 & 37.71 & 57.00\\\hline
hugetrace-00020 & {\bf 7.90} & 393.13 & 15.95 & 15.02 & {\bf 1.52} & 9.97 & 8.68 & 38.27\\\hline
hugebubbles-00000 & 13.16 & {\bf 3.55} & 19.81 & 5.56 & {\bf 1.80} & 10.91 & 10.03 & 38.97\\\hline
wb-edu & 33.82 & 8.61 & {\bf 3.38} & 20.35 & 17.43 & 20.10 & {\bf 9.49} & 51.14\\\hline
rgg\_n\_2\_24\_s0 & 3.68 & 2.25 & 25.40 & {\bf 0.12} & {\bf 2.20} & 12.50 & 5.72 & 31.78\\\hline
patents & 0.88 & {\bf 0.84} & 92.03 & 16.18 & {\bf 0.91} & 0.97 & 101.76 & 18.30\\\hline
italy\_osm & 5.86 & 1.20 & {\bf 1.02} & 122.00 & {\bf 0.70} & 3.97 & 6.24 & 18.34\\\hline
soc-LiveJournal1 & {\bf 3.32} & 14.35 & 243.91 & 21.16 & {\bf 3.73} & 7.14 & 343.94 & 20.71\\\hline
ljournal-2008 & {\bf 2.37} & 10.30 & 360.31 & 17.66 & {\bf 6.90} & 7.58 & 176.69 & 23.45\\\hline
europe\_osm & 57.53 & {\bf 11.21} & 14.15 & 1911.56 & {\bf 7.21} & 37.93 & 68.18 & 197.03\\\hline
com-livejournal & {\bf 4.58} & 22.46 & 2879.36 & 34.28 & {\bf 5.88} & 17.19 & 165.32 & 29.40\\\hline

\end{tabular}
}
\end{table}

\section{Concluding remarks}\label{sec:con}

We proposed a parallel BFS based GPU implementation of maximum cardinality 
matching algorithm for bipartite graphs. We presented experiments on various 
datasets, and compared the performance of the proposed GPU implementation against sequential and multicore algorithms. The 
experiments showed that the proposed GPU implementations are faster than the 
existing parallel multicore implementations. 
The speedups achieved with respect to well-known sequential implementations varied from $0.03$ to $629.19$, averaging $9.29$ on a set of 20 hardest problems with respect to the fastest sequential algorithm.
A GPU is a restricted memory device. 
Although, an out-of-core or distributed-memory type algorithm is amenable when the graph does not fit into the 
device, a direct implementation of these algorithms will surely not be efficient. 
We plan to investigate the techniques to obtain good matching performance for extreme-scale bipartite graphs on GPUs. 

\vspace*{-2ex}
\bibliographystyle{splncs03}

\end{document}